# Harnessing transformation optics for understanding electron energy loss and cathodoluminescence

Yu Luo [1, 2] *, Matthias Kraft [1] *, and J. B. Pendry[1]

[1] The Blackett Laboratory, Department of Physics, Imperial College London, London SW7 2AZ, United Kingdom

[2] School of Electrical and Electronic Engineering, Nanyang Technological University, Nanyang Avenue 639798, Singapore

* These authors contributed equally

**Abstract**

As the continual experimental advances made in Electron energy loss spectroscopy (EELS) and cathodoluminescence (CL) open the door to practical exploitations of plasmonic effects in metal nanoparticles, there is an increasing need for precise interpretation and guidance of such experiments. Numerical simulations are available but lack physical insight, while traditional analytical approaches are rare and limited to studying specific, simple structures. Here, we propose a versatile and efficient method based on transformation optics which can fully characterize and model the EELS problems of nanoparticles of complex geometries. Detailed discussions are given on 2D and 3D nanoparticle dimers, where the frequency and time domain responses under electron beam excitations are derived.

**Significance Statements**

Critical to the wide applications of nanoplasmonics is the ability to probe the near-field electromagnetic interactions associated with the nanoparticles. Electron microscope technology has been developed to image materials at the subnanometer level, leading to experimental and theoretical breakthroughs in the field of plasmonics and beyond. This has brought increasing need for precise interpretation and guidance of such experiments. To address this concern, we propose a versatile method based on transformation optics which can fully characterize and model the electron energy loss problems for nanoparticles interacting with electron beams. Our methodology gives rise to an accurate and extremely efficient computational scheme that will be of value in further studies of nanostructures, yielding unprecedented insight into the optical responses in nanoplasmonic systems.

## Introduction

The noble metal nanoparticles have received considerable attention because they exhibit localized surface plasmon resonances (LSPRs), where the collective excitations of the conduction electrons couple to light and can compress the captured energy into just a few cubic nanometers (1). Frequency and intensity of the resonances are dictated by the nanoparticle's material, size, shape, electronic charge and surrounding medium (2, 3). Controlling the geometry of the nanoparticles to produce the desired LSP modes allows for various applications of nanoplasmonics ranging from biosensing, gas molecule detection, fluorescence, and thin-film photovoltaics, to optical data storage and on-chip wave guiding (4). Critical to these applications is the ability to probe the near-field electromagnetic interactions associated with the nanoparticles; a thorough understanding of the plasmonic response is paramount for the design of nanostructures with tailored optical properties. However, surface plasmon phenomena occur at subwavelength scales not observable with conventional diffraction-limited optical microscopes. Electron microscopy technology has overcome this limitation and thus led to experimental and theoretical breakthroughs in the field of plasmonics and beyond (5-11). One, early, variant of electron microscopy is the study of plasmons in nanoparticles, by detecting the photons emitted under electron irradiation (cathodoluminescence spectroscopy, CL) in a scanning transmission electron microscope (STEM) (12-15). In recent years, electron energy-loss spectroscopy (EELS) has grabbed a lot of attention in the plasmonics community. Its extremely high spatial resolution (sub-nanometer) and very good energy resolution (below 100 milli eV) make it ideally suited to probe plasmon modes in nanoparticles in a broad spectral range, from the infrared to ultraviolet (IR-vis-UV) frequencies (16-20). Moreover, electron beams can excite dark modes, which have a zero net dipole moment and are not accessible with far field light-based measurement techniques (21, 22). These combined advantages make EELS an ideal tool in spectral and direct morphological analysis of individual nanoparticles, particularly for nanoparticles with features on the sub-nanometer scale, where plasmon resonances become sensitive to quantum effects of the conduction electrons (23-25). In this small size regime, EELS provides the key to resolving many debates that were due to the lack of effective probing methods.

The continual experimental advances made in EELS open the door to practical exploitations of plasmonic effects. However, they also increase the need for interpreting and guiding experiments. Numerical simulations are widely used to support experiments (26-28) but provide limited physical insight. Analytical approaches based on multiple scattering theory have been adopted to study isolated spheres (29, 30) or clusters of spheres (31, 32); the results were confirmed in EELS experiments (33-36). These investigations, however, are limited to specific, simple structures. Moreover, as the multiple scattering method is very demanding in terms of computational resources, the time-domain response of the system, which is important for understanding EELS experiments, has never been obtained. A versatile and efficient method which can fully characterize and model EELS problems of complex plasmonic systems is still in demand. Transformation optics (TO) provides an elegant solution.

TO has emerged over the last two decades, as a powerful technique for the control of electromagnetic (EM) fields (37-40). An intuitive to use tool that facilitates system design and analysis, it is yet fully rigorous at the level of Maxwell's equations. Owing to these advantages, TO offers a unique way to studying subwavelength EM phenomenon and has already been applied successfully to analyze rigorously and design systematically plasmonic nanostructures for different applications (41-44). Here, we show that TO can also serve as a powerful tool in understanding EELS problems for complex plasmonic nanoparticles. As an example, we consider two closely spaced spherical and cylindrical dimers. Using TO, we derive analytical expressions for the electron energy loss probability and photon scattering spectra, and calculate the electric fields in the frequency domain. The analytical nature of our approach means we can also access the time evolution of the system, which is difficult to obtain with traditional methods, in a computationally efficient manner by a simple fast Fourier transform. Note that our investigation can be easily extended to treat EELS problems in other sharp tip or small gap geometries (see Ref. (41) and references therein), but the dimer system is interesting for several reasons. Closely spaced nanoparticles can interact, that is, their plasmon modes hybridize, which gives rise to a much richer plasmon spectrum than that of the isolated particles (45). Such a spectrum does not only include bright modes but also dark modes (with vanishing dipole moments and radiation loss) that cannot be probed by optical absorption experiments, but can with EELS. The strong interactions between the nanoparticles also result in highly localized electric fields, leading to a tremendous enhancement in the optical energy density in the gap between nanoparticles (46).

The paper is structured as follows. We outline the Transformation optics approach and give a Transformation from closely spaced spheres/nanowires to a concentric annulus and also show how the electron trajectories transform. We then discuss the frequency and time domain response of the 2-D nanowire dimer system. Finally, we provide an in depth analysis of the 3-D spherical dimer system, discussing its photon emission spectrum, its electron energy loss spectrum, their integrated version and, last but not least, analyze the time-domain response of the 3-D dimer.

**Transformation of the geometry**

The two sphere/cylinder systems are the simplest strongly interacting plasmonic systems and thus ideal to test our method, as shown in Figure 1 (a). TO simplifies the study of a system with complex geometry by transforming it into a highly symmetric virtual system, where the optical responses can be easily solved. In a standard TO approach, all the expressions for the optical response in this transformed system (virtual frame) are then transformed back to the physical frame to find the response of the desired system. Standard transformation rules for quantities of interest exist (38), the most important for our present study is the invariance of the electrostatic potential, though.

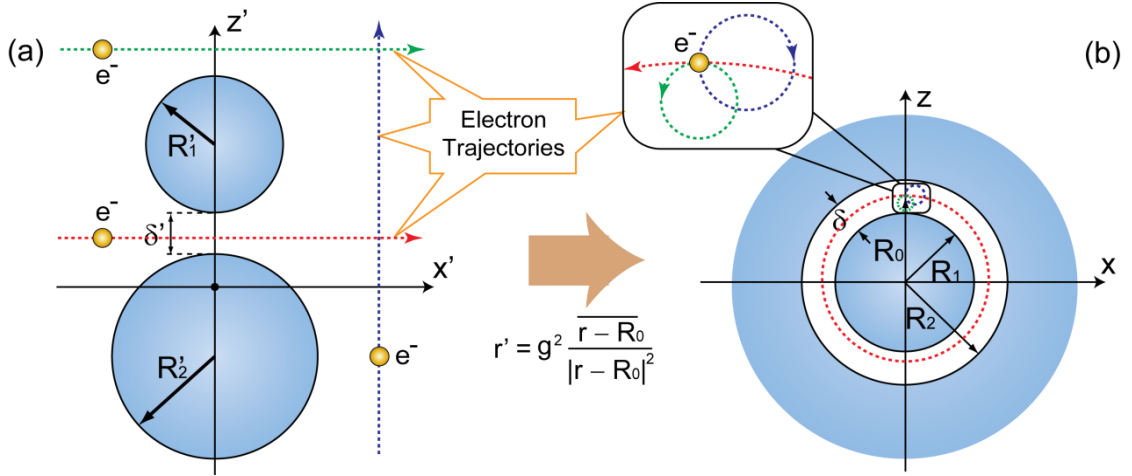

**Figure 1 Transformation optics enables a simple calculation of the electron energy loss problem.** The schematic of the 2-D/3-D transformation that relates a pair of closely spaced nanowires/spheres (a) to a concentric annulus (b), whose analytical solution can be obtained easily. The dashed lines represent the trajectories of the moving electron. In the original space (a), the electron travels along a straight line from minus to positive infinity at a uniform speed. In the transformed geometry (b), the electron moves in a circle with a non-uniform velocity, starting from and then returning to the inversion point, as illustrated in the top left inset.

Here we need to differentiate between the 2-D and 3-D cases. In both cases, we can transform the cylindrical/spherical dimer to a 2-D/3-D annulus. In 2-D, this is achieved using a conformal transformation, which leaves the material parameters unchanged under the transformation. In 3-D, we apply an inverse transformation, which means the 3-D annulus acquires a space dependent permittivity, but that need not worry us (46). We are interested in the response of this system when an electron moves past the dimers horizontally outside the gap, horizontally inside the gap and vertically outside the gap (see Figure 1a). We assume the electron moves at a constant speed. Since charge is conserved under coordinate transformations the electron trajectory in the transformed frame is dictated by the coordinate transformation, yielding three circular trajectories (see Figure 1b). However, as space is compressed non-uniformly by this transformation, the electrons in the transformed frame possess a space dependent velocity, starting slowly at the origin, moving more rapidly at the farthest point from the origin, and slowing down again before returning to the origin. This makes it difficult to calculate the electric fields associated with them. Instead we propose the following solution method: the electric field associated with the moving electron in free space is first calculated in the physical frame (Figure 1a). It is then transformed to the virtual frame using the invariance of the electrostatic potential. We then expand that potential in terms of the plasmon eigenmodes of the 2-D/3-D annulus and match source and scattered potentials using the standard boundary conditions on the electric fields. The solution in the physical (dimer) frame is then obtained by transforming the solution back from the virtual frame. This unambiguously determines the response of the dimer system and allows us to calculate: the electric fields in the frequency domain, the total electron energy loss, the photon scattering spectrum and the electric fields in the time domain via a Fourier transform. The mathematical details are provided in the supplementary materials; here we focus mainly on the results and the physical interpretations.

**Results: 2-D dimer**

In EELS and CL experiments, a beam of monoenergetic fast moving electrons interacts with the plasmonic nanoparticles. The electric field associated with the fast flying electrons affects the free electrons in the metal nanostructures and generates a charge displacement, leading to a collective oscillation of conduction electrons at their plasmon resonance frequency and emission of electromagnetic radiation. Due to the absorption and scattering, the electrons are losing the respective amount of energy, which is, however, negligible compared to their total energy. The electron energy loss spectrum of the electron beam after passing the nanostructure provides information on the plasmonic modes. Of course, in 2-D we are really dealing with a line electron, so all the results are per unit length (see also Ref (47)).

Figure 2 (a1)/(b1) shows the spectrum of an electron moving outside the gap of the cylindrical dimer. In both cases, the strength of the plasmon resonances monotonically increases with the electron's velocity in this velocity range. This is in contrast to Figure (c1) where the electron passes through the centre of the gap between the two dimers. Here, the slowest electron (black line) actually leads to the strongest excitations. Additionally, only modes above the surface plasma frequency are excited. This is due to the symmetry properties of the dimer as the modes above the surface plasma frequency have an even potential with respect to reflections about the axis through the centre of the gap. When the electron passes exactly through the middle of the gap its potential has exactly the same symmetry, meaning only the modes above the surface plasma frequency are excited. Modes below the surface plasma frequency can be excited only if the electron passes through the gap 'off' centre. Figure 2 (a2-c4) shows the time response of the dimers, as we plot the vertical component of the electric fields at 0fs, 4fs and 8fs after the electron passes the vertical axis of the system. Both excitation and the subsequent decay of the surface plasmons can be observed. Most intriguingly though, it can be seen how the system 'harvests' energy and concentrates it at a hotspot. For the top and middle row, plasmons are excited at the outer edge of the dimer structure, but are then squeezed into the gap between the cylinders. This is most easily seen from the video simulations provided as supplementary material.

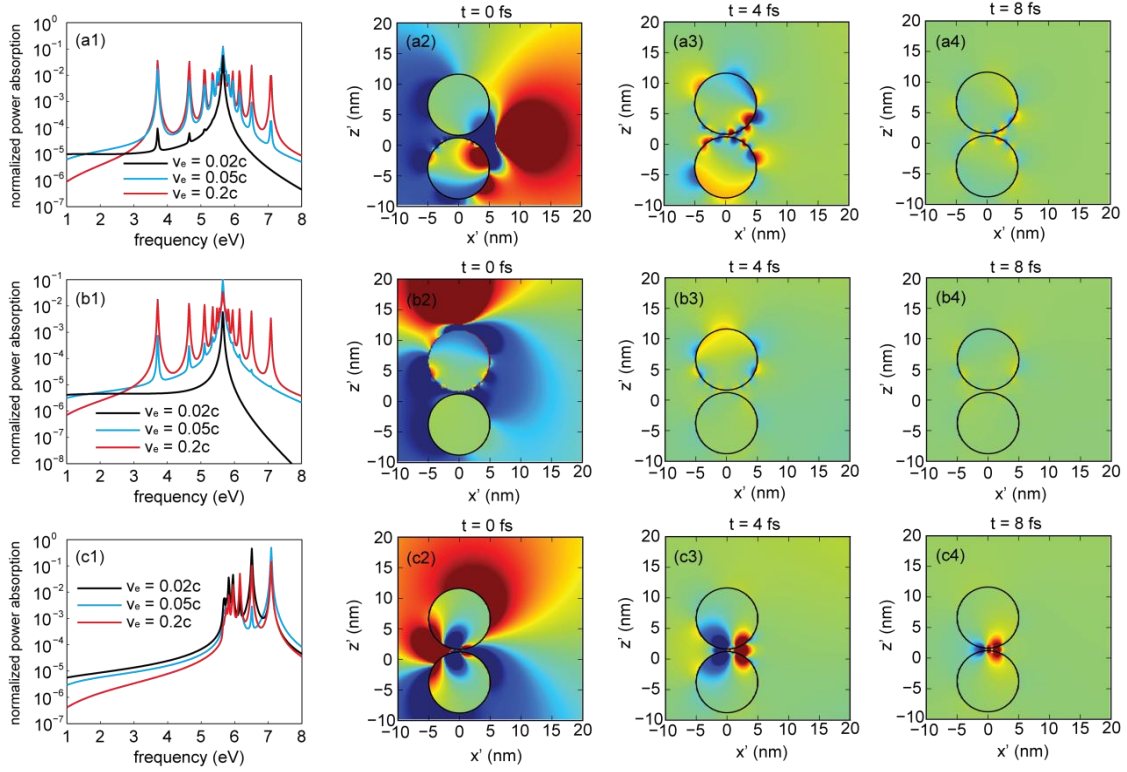

**Figure 2 Electron energy loss for cylindrical dimer.** Electron energy loss spectra (a1)-(c1) and corresponding time-domain field ($E_x$) distributions at three different time points $t = 0$ fs (a2)-(c2), 4 fs (a3)-(c3), and 8 fs (a4)-(d4). The top panels illustrate the cases where the electron travels along the vertical $z'$ direction; the middle panels correspond to the electron moving along horizontal $x'$ direction above the two nanowires; the bottom panels represent the cases where the electron moves across the gap. The radii of the two nanowires are set as $R'_1 = R'_2 = 5$ nm. The separation between them is $\delta' = 0.4$ nm. The distance from the nanoparticle to the electron trajectory is 0.2nm. The metal permittivity is given by the Drude model $\varepsilon_m = 1 - \omega_p^2 / \left[ \omega(\omega + i\gamma) \right]$ with $\omega_p = 8$ eV and $\gamma = 0.032$ eV. Curves of different colors in (a1)-(c1) correspond to different electron velocity. The time domain field distributions are calculated for $v_e = 0.05c$.

**Results: 3-D dimer**

The 2-D calculations capture the response to electrons which lose momentum only in the $x'$-$z'$ plane. In reality, however, an electron beam passing the nanoparticle excites plasmon modes in all three dimensions, and hence loses energy/momentum in both the in-plane and out-of-plane directions. Thus, to compare with experiments for EELS and CL spectra of two closely spaced cylinders, one has to filter out the electrons that lost momentum in the $y'$-direction. To capture the complete response of a plasmonic system we need to extend our approach to 3-D. We thus consider a 3-D system next, which comprises two nearly touching spheres under electron bombardment under different trajectories, as shown in the leftmost column of Figure 3. The extension from 2-D to 3-D is nontrivial. A summary of the derivations is given below and details can be found in the supplementary material.

Solving the Laplace equation in the transformed annulus geometry, the electrostatic potential can be written in spherical harmonics as:

$$\begin{aligned}
\phi &= a_{lm}^{in} |r-R_0| (r/R_0)^l Y_{lm}(\theta,\varphi), & r &< R_1, \\
&= |r-R_0| \left[ \left( a_{lm}^+ + a_{lm}^{s+} + a_{lm}^{r+} \right)(r/R_0)^l + a_{lm}^- (r/R_0)^{-(l+1)} \right] Y_{lm}(\theta,\varphi), & R_0 &> r > R_1, \\
&= |r-R_0| \left[ a_{lm}^+ (r/R_0)^l + \left( a_{lm}^- + a_{lm}^{s-} + a_{lm}^{r-} \right)(r/R_0)^{-(l+1)} \right] Y_{lm}(\theta,\varphi), & R_2 &> r > R_0, \\
&= a_{lm}^{out} |r-R_0| (r/R_0)^{-(l+1)} Y_{lm}(\theta,\varphi), & r &> R_2.
\end{aligned} \qquad (1)$$

The complex expansion coefficients $a_{lm}^{s+}$ and $a_{lm}^{s-}$ associated with the moving electron are derived in the supplementary materials. Applying the boundary conditions and solving the tridiagonal reflection matrix (46) gives solutions for the unknown expansion coefficients $a_{lm}^+$, $a_{lm}^-$, $a_{lm}^{r+}$, $a_{lm}^{r-}$, $a_{lm}^{in}$ and $a_{lm}^{out}$. Detailed derivations given in the supplementary materials show that the power associated with the photon emission is calculated as:

$$P^{sca} = \frac{c\varepsilon_0 k_0^4 g^8}{3R_0^2} \left\{ \left[ \sum_{l=0}^{\infty} \sqrt{2l+1} \left( l a_{l0}^+ - (l+1) a_{l0}^- \right) \right]^2 + \left[ \sum_{l=1}^{\infty} \sqrt{l(l+1)(2l+1)} \left( a_{l1}^+ + a_{l1}^- \right) \right]^2 \right\}, \qquad (2)$$

while electron energy loss takes the form of

$$P^{el} = -\frac{q}{4\pi} \operatorname{Im} \left[ \sum_{m=-\infty}^{+\infty} \sum_{l=|m|}^{\infty} \left( s_{lm}^+ a_{lm}^+ + s_{lm}^- a_{lm}^- \right) \right], \qquad (3)$$

where $s_{lm}^+$ and $s_{lm}^-$ are constants depending on the electron trajectory:

$$s_{l,m}^+ = \begin{cases} (-1)^m \dfrac{2(2l+1)\pi\omega\varepsilon_0 R_0 g^2 e^{-i\omega g^2/(vR_0)}}{q} a_{lm}^{s\pm}, & \text{for vertical case} \\[2mm] (-1)^m \dfrac{2(2l+1)\pi\omega\varepsilon_0 R_0 g^2}{q} a_{lm}^{s\mp}, & \text{for horizontal case} \\[2mm] (\mp 1)^m \dfrac{2(2l+1)\pi\omega\varepsilon_0 R_0 g^2}{q} a_{l,m}^{s\mp}, & \text{for gap case} \end{cases}$$

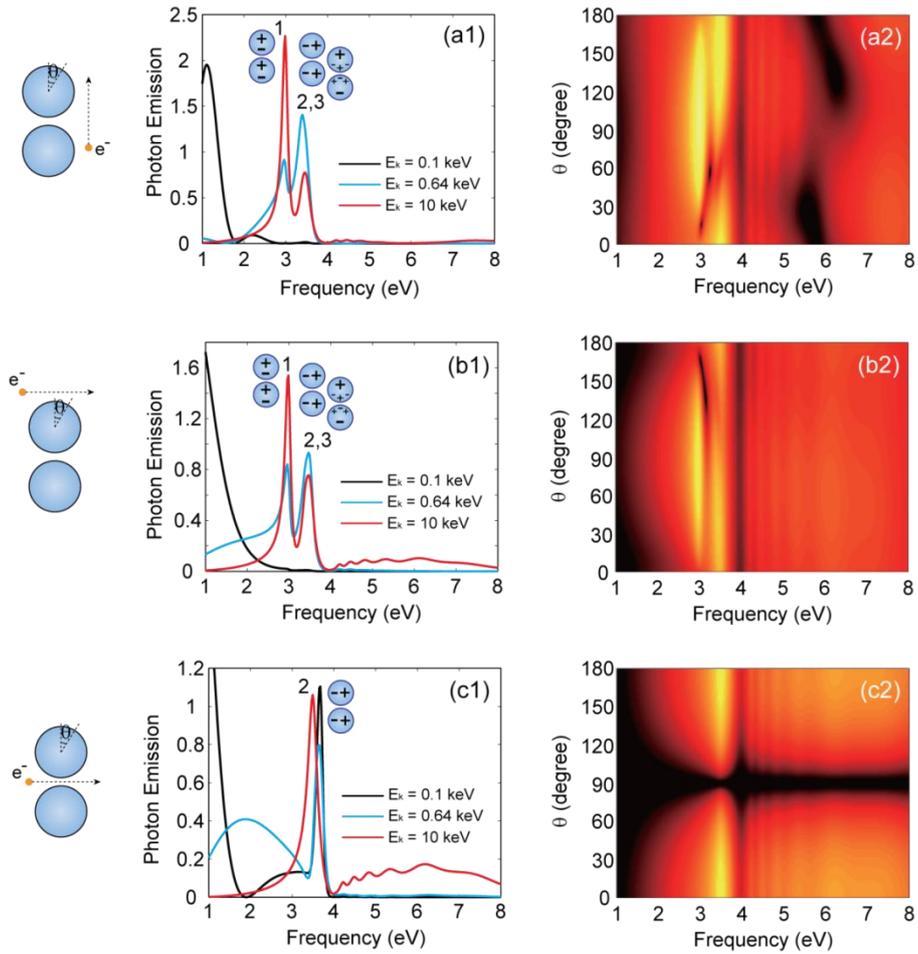

**Figure 3 Photon emission for spherical nanoparticle dimer.** (a1)-(c1) Photon emission spectra; (a2)-(c2) Contour plots of the photon emission pattern as a function of the frequency and the polar angle $\theta$. Panels from top to bottom correspond to the electron moving along the vertical direction, the horizontal direction alongside the two spheres, and the horizontal direction across the gap between the two spheres, respectively. The radii of the two spheres are chosen as $R_1' = R_2' = 10$ nm. The separation between them is $\delta' = 0.8$ nm. The distance from the electron trajectory to the closest sphere is set as 0.4nm for all three cases. Curves of different colors in (a1)-(c1) correspond to different electron kinetic energy. The scattering patterns in (a2)-(c2) are calculated for $E_k = 10$ keV. We used experimental data permittvity data of silver (48) for the dimer and assumed the surrounding medium to be vacuum.

To demonstrate our methodology the above formulae are applied to calculate the response of two identical spheres of radius $R_1' = R_2' = 10\,\text{nm}$ separated by $\delta' = 0.8\,\text{nm}$. Experimental data for the permittivity of silver is used in the calculation for the closely spaced spheres (48). In Figures 3 (a1-c1), the emitted photon number spectra are compared for different kinetic energies ($E_k$=0.1keV, 0.64keV, 10keV) of the electron. As denoted by the black curve, an electron moving vertically (Figure 3 (a1)) or horizontally (Figure 3 (b1)) at the edge of spherical dimer with very small kinetic energy ($E_k$=0.1keV) cannot excite any plasmon modes. As the electron energy increases to $E_k$=0.64keV, three dipole-active bright modes can be observed, independently of the direction of the moving electron. The second and third modes overlap and a detailed analysis requires a discussion of the system's time domain response, see Figure 6 and its explanation. For both vertical and horizontal edge excitation cases, further increasing the electron's velocity significantly enhances the excitation of the first dipolar mode, while modes 2 and 3 are less sensitive to the increase. When the electron is moving horizontally across the center of the gap, as shown in Figure 3 (c1) and (c2), only mode 2 can be excited due to the symmetry properties of the dimer. Similar to the 2D case, the change in the electron's energy only results in small variation of this mode and the photon emission is strong even for very small energies around $E_k$=0.1keV. The right column displays the contour plots of the photon emission pattern as a function of frequency and polar angle $\theta$ for different electron trajectories at kinetic energy $E_k = 10\,\text{keV}$, showing that mode 1 radiates along the horizontal direction while mode 2 radiates in the vertical direction.

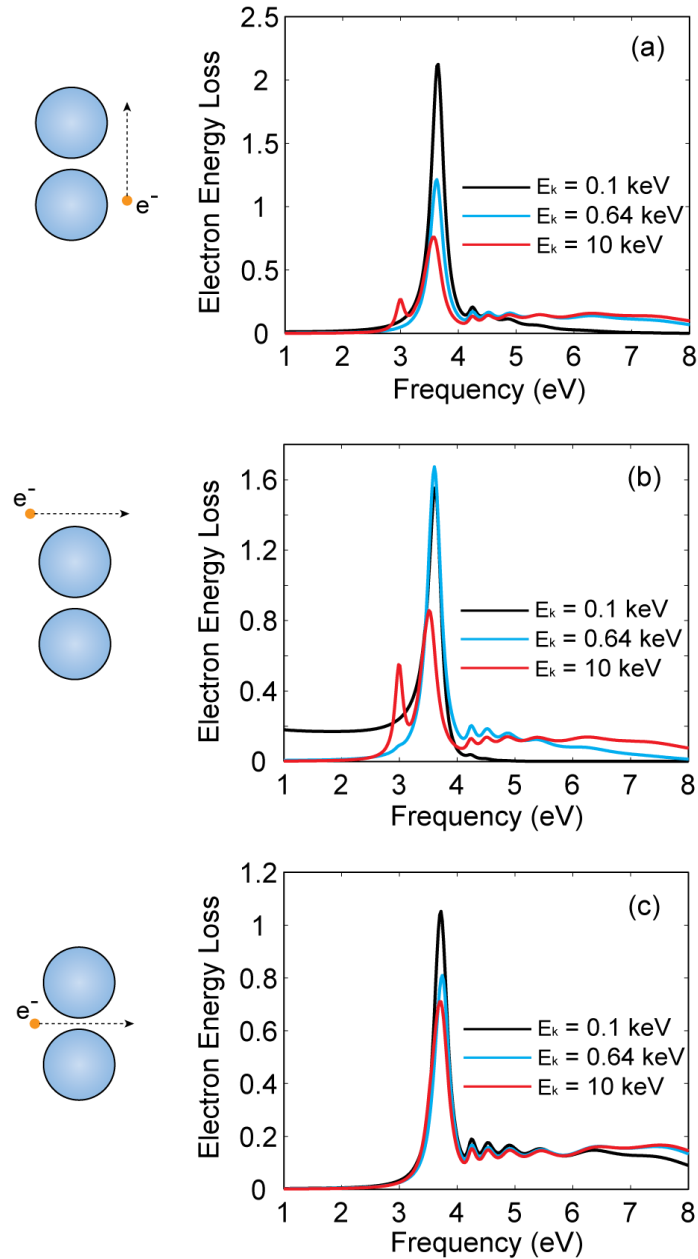

**Figure 4 Electron energy loss for spherical nanoparticle dimer.** The energy loss for the electron travelling along (a) the vertical direction; (b) the horizontal direction alongside the two spheres; (c) the horizontal direction across the gap between the two spheres. We used the same geometrical and permittivity parameters as in Figure 3.

In contrast to CL which probes only the radiative modes, EELS measures the total electron energy loss, revealing additional modes which have vanishing dipole moments. Figure 4 depicts the electron energy loss spectra for the same setup as in Figure 3. For electrons moving outside the gap with kinetic energy $E_k$=0.1keV no modes were present in the photon emission spectra around 3.5eV. Yet, the EELS spectra in Figure 4a and 4b feature a strong resonance at 3.5eV for this kinetic energy, indicating the presence of dark modes. Increasing the electron energy gives rise to another peak at 3eV, corresponding to the first dipolar mode as observed in Figure 3, as well as a complex peak around 3.5eV comprising two dipolar modes overlapping with several dark modes in the spectrum. Similar phenomena can be observed in the gap excitation case except for the absence of the first dipolar mode.

An independent analysis of each dark mode seems hopeless in this case, as they all resonate in a very small frequency window. However, this limitation can be overcome by covering the silver spheres with a dielectric coating, which will spread out the collective dark modes in the spectrum, allowing each mode to be clearly probed (Detailed analysis is given in supplementary documents).

Figure 5 gives the dependence of total photon emission and total electron energy loss on the kinetic energy of moving electron. For all three excitations, the total photon emission grows monotonically with the electron's energy. However, contrary to what one might expect, the total electron energy loss is not a monotonic function of the electron velocity. For excitation from outside the gap, the total energy loss reaches its maximum at around 1keV and plateaus until 10keV. For excitation through the gap, the total energy loss remains stable at a high level even for very small electron energies (milli-keVs). This result indicates that a fast moving electron with kinetic energy tens of keV (as commonly used in EELS experiment) is not always desired to excite specific modes.

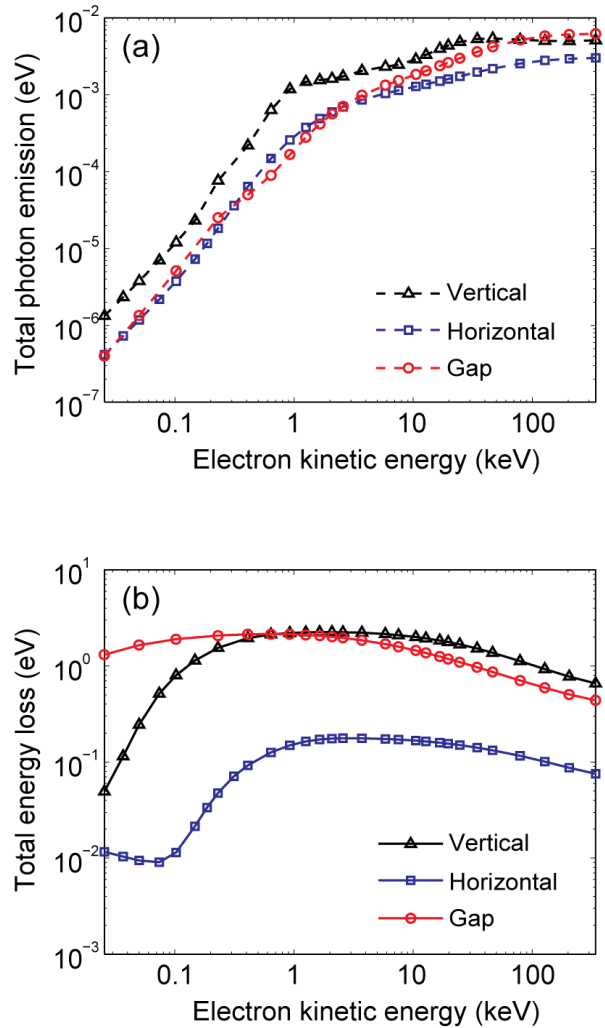

**Figure 5 Total photon scattering and total energy loss.** The total (integrated) energy of photon scattering (a) and the total energy of electron loss (b) as a function of the electron kinetic energy. We used the same geometrical and permittivity parameters as in Figure 3 and 4.

Thanks to the analytical and computationally efficient nature of the TO approach, we can easily obtain the time response of the sphere dimer by applying a Fourier transform to the potential. Compared to traditional methods which require immense computer resources to calculate the time response, our calculations can be performed within tens of minutes on a standard laptop using code implemented in Mathematica. Figure 6 shows the time evolution of the electric field as electrons move past the spherical dimer. We plot the normalized *x* and *z* components of the electric field at the center of the gap for three different electron trajectories and kinetic energies. The insets in Figure 6 show the corresponding field enhancements in the frequency domain. The fields as a function of time allow us to distinguish the contributions from antibonding and bonding modes to the spectra, due to the symmetry properties of the structure (a detailed analysis is given in the supplementary materials). We find that independently of the electron's trajectory, the $E_x$ response (red curve) is dominated by an antibonding dipolar mode at 3.5eV. Further, the electron moving across the center of the gap exclusively excites this antibonding mode, with the smallest velocity yielding the strongest excitation (Panel (c1)). In contrast, electrons moving outside the gap can excite bonding modes, too (panel (a2-3, b2-3)), but need considerable energy to do so, as indicated by the absence of modes in panels (a1) and (b1). This effect is most easily seen from the z-component of the electric field. For a vertically moving electron at moderate velocity, the $E_z$ response is dominated by the bonding dipolar and bonding quadrupole modes, where the latter overlaps with the antibonding dipolar mode at 3.5eV, as depicted by Figure 6 (a2). Note that these two modes with the same spectral position correspond to mode 2 and 3 in the photon emission spectra in Figure 3. In all the other cases, the $E_z$ response is dominated by the bonding dipolar mode, which is most strongly excited by the vertically moving electron with the largest velocity.

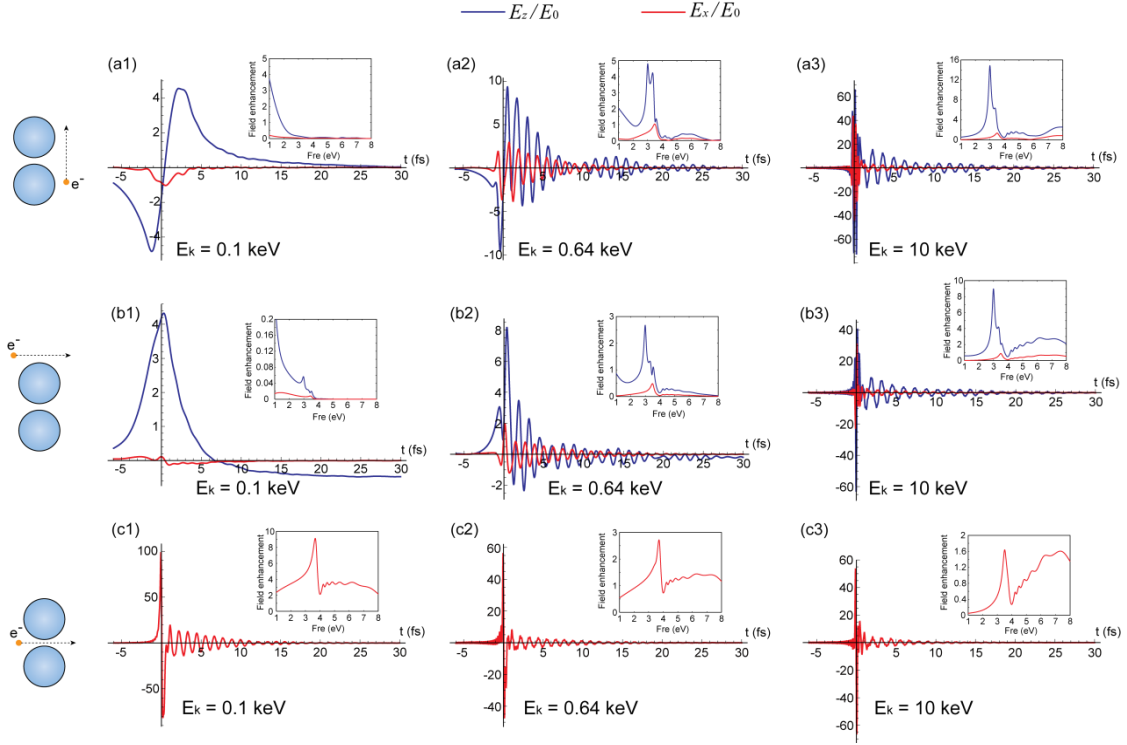

**Figure 6 Time domain field response.** The normalized electric fields $E_x/E_0$ (red) and $E_z/E_0$ (blue) at the center of the gap as a function of time. $E_0$ is the magnitude of the electric field at the same position associated with an electron moving on the same trajectory in free space without the dimer structure. The inset in each panel shows the field enhancements in the frequency domain. Different panels correspond to the electron travelling on different trajectories and with different velocities. The geometrical parameters of the spherical dimer are the same as in Figure 3. The metal permittivity at the frequency range where experimental data are not available is obtained by fitting (49).

**Conclusion**

To conclude, we have presented a new TO-based approach to calculating the electron energy loss and photon emission spectra, as well as the time domain responses for 2D and 3D nanoparticles under the electron beam excitations. Specifically, we considered two closely spaced spherical and cylindrical dimers, representing the most fundamental system of two interacting objects. Our quantitative analysis indicates that the total energy loss is not a monotonic function of the electron velocity and hints at the optimal velocity and injection position to maximize the excitation of the modes. Our methodology gives rise to an accurate and extremely efficient computational scheme that will be of value in further studies of plasmonic nanostructures. The analytical studies provide important guidance and precise interpretation for EELS and CL experiments, yielding unprecedented insight into the plasmonic responses in this and related systems.

**Acknowledgement**

JBP acknowledges support from the EPRSC Mathematical Fundamentals of Metamaterials Programme (EP/L024926/1) and from the Gordon and Betty Moore Foundation; MK acknowledges support from the Imperial College PhD scholarship; YL acknowledges support from Singapore Ministry of Education under Grant No. TIER 1 (RG72/15) and Grant No. MOE2015-T2-1-145.